\documentclass[sigconf]{acmart}

\usepackage{tikz}
\usetikzlibrary{arrows.meta,positioning,fit,backgrounds,calc}
\definecolor{FigNavy}{HTML}{344256}
\definecolor{FigMuted}{HTML}{64748B}
\definecolor{FigLavender}{HTML}{EEF2FF}
\definecolor{FigBlue}{HTML}{E2F2FC}
\definecolor{FigMint}{HTML}{E8F7EF}
\definecolor{FigCream}{HTML}{FFF4E8}
\definecolor{FigTab}{HTML}{0E7490}
\definecolor{FigLine}{HTML}{B9C2CC}
\tikzset{
  figbase/.style={draw=FigNavy, line width=0.7pt, rounded corners=2pt,
    align=left, inner sep=4pt, font=\sffamily\tiny, text=FigNavy},
  figinput/.style={figbase, fill=FigLavender},
  figinterface/.style={figbase, fill=FigBlue},
  figfocus/.style={figbase, fill=FigBlue, draw=FigTab, line width=1.2pt},
  figprocess/.style={figbase, fill=FigMint},
  figoutput/.style={figbase, fill=FigCream},
  figchip/.style={draw=FigLine, fill=white, rounded corners=1.5pt,
    align=center, inner sep=2.5pt, font=\sffamily\tiny, text=FigNavy},
  figure-readable/.style={every node/.append style={font=\sffamily\scriptsize}},
  figarrow/.style={-{Latex[length=1.8mm]}, draw=FigNavy, line width=0.7pt},
  figdash/.style={-{Latex[length=1.8mm]}, draw=FigTab, dashed, line width=0.7pt}
}
\setcopyright{none}
\renewcommand\footnotetextcopyrightpermission[1]{}
\acmConference[RecSys Challenge '26]{Proceedings of the Recommender Systems Challenge 2026}{September 28--October 2, 2026}{Minneapolis, MN, USA}
\acmYear{2026}
\copyrightyear{2026}

\begin{document}

\title[Two Views, One Voice]
{Two Views, One Voice: Evidence-Grounded Conversational Music Recommendation}

\author{Sungwook Yoo}
\authornote{Both authors contributed equally to this research.}
\email{l22491360@gmail.com}
\affiliation{\institution{Naver}\country{South Korea}}

\author{Sewook Yoo}
\authornotemark[1]
\email{sewookyoo@gmail.com}
\affiliation{\institution{Samsung}\country{South Korea}}

\begin{abstract}
Traditional conversational recommenders entangle retrieval and response generation within a single text interface, so exact entity cues fade as the dialogue's intent evolves, which compromises explanation credibility. We address this within the ACM RecSys Challenge 2026, which mandates both top-20 ranking and evidence-grounded response generation. This paper presents the third-place solution by team ``swyoo'' for the Blind-B industry track. We decouple retrieval and response into separate pipelines connected strictly via ranked tracks and metadata. Retrieval combines a hybrid lexical-dense pool for exact matching with a task-adapted pool driven by fine-tuned Qwen 8B adapters. Candidates are calibrated via LightGBM, then routed to an evidence-grounded propose-assign-select (PAS) framework to structure responses. This system also ranked second on the explanation-quality leaderboard in the final blind evaluation. Our findings demonstrate that: (i) isolating retrieval and response preserves both catalog cues and fluid intent; (ii) structuring generation via explicit evidence assignment is key to this near-best-in-class explanation reliability.
\end{abstract}



\maketitle

\section{Introduction}

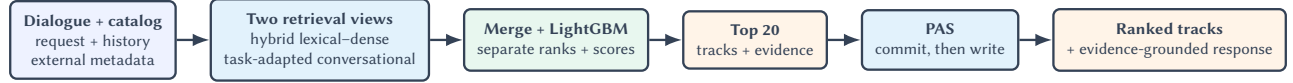
\begin{figure*}[t]
  \centering
  \resizebox{0.95\textwidth}{!}{%
  \begin{tikzpicture}[node distance=3.5mm,figure-readable]
    \node[figinput,align=center] (input)
      {\textbf{Dialogue + catalog}\\request + history\\external metadata};
    \node[figinterface,align=center,right=3.5mm of input] (pools)
      {\textbf{Two retrieval views}\\
       hybrid lexical--dense\\
       task-adapted conversational};
    \node[figprocess,align=center,right=3.5mm of pools] (merge)
      {\textbf{Merge + LightGBM}\\separate ranks + scores};
    \node[figoutput,align=center,right=3.5mm of merge] (top)
      {\textbf{Top 20}\\tracks + evidence};
    \node[figinterface,align=center,right=3.5mm of top] (pas)
      {\textbf{PAS}\\commit, then write};
    \node[figoutput,align=center,right=3.5mm of pas] (out)
      {\textbf{Ranked tracks}\\+ evidence-grounded response};

    \draw[figarrow] (input.east) -- (pools.west);
    \draw[figarrow] (pools.east) -- (merge.west);
    \draw[figarrow] (merge.east) -- (top.west);
    \draw[figarrow] (top.east) -- (pas.west);
    \draw[figarrow] (pas.east) -- (out.west);
  \end{tikzpicture}}
  \caption{System overview. Two retrieval views merge only at calibration;
  PAS grounds the response in the fixed top 20.}
  \Description{Conversation and externally enriched catalog inputs feed the bm25 qmr and qemb two-tower 8B pools. Their candidates are merged and calibrated by LightGBM, after which the top 20 and evidence flow through PAS to the ranked tracks and response.}
  \label{fig:pipeline}
  \vspace{-2mm}
\end{figure*}

Conversational music recommendation (CRS) must balance precise catalog entity retrieval with fluid dialogue tracking~\cite{melchiorre2025jam,kemper2024rarec}. Traditional architectures route both tasks through a single text interface, so as later turns emphasize evolving preferences, earlier entity mentions are diluted within the accumulated query text and dropped from retrieval, undermining the exact catalog matches that credible explanations require~\cite{qin2024credible}. To resolve this, we propose an operationally decoupled architecture that separates retrieval and response. As shown in Figure~\ref{fig:pipeline}, these pipelines connect strictly via ranked candidates and metadata evidence, directly aligning with the ACM RecSys Challenge 2026\footnote{\url{https://www.recsyschallenge.com/2026/}} objectives of top-20 ranking and evidence-grounded response generation.

Figure~\ref{fig:textviews}b illustrates this single-interface vulnerability. At turn 4, a listener requests ``melancholic darkwave'' citing \textit{Switchblade Symphony}; by turn 6, they add ``gothic rock'' and a ``heavier beat'' without repeating the artist. A keyword baseline (\texttt{bm25\_qmr}) drops the ground-truth track---``Dollhouse (Razed In Black Mix)''---to 20th place due to this entity shift. Our system instead runs a hybrid lexical--dense pool for exact matching alongside a conversational pool driven by fine-tuned Qwen 8B adapters (Figure~\ref{fig:textviews}a): the 8B pool alone recovers the target to fourth place, RRF of the two views lifts it to second, and LightGBM ranks it first.

After ranking, generating explanations requires strict factual grounding. Instead of relying on free-form generation, we adapt an evidence-grounded PAS framework (Figure~\ref{fig:pas}). PAS converts response generation into a joint reasoning step conditioned on candidate-level evidence and a matched demonstration, followed by deterministic validation to anchor outputs to verified catalog attributes. We release our full pipeline\footnote{\url{https://github.com/yoobros/music-crs-challenge}}. Our main technical contributions are:
\begin{itemize}
    \item An operationally decoupled CRS framework separating retrieval from response generation.
    \item A multi-view calibration pipeline merging hybrid and Qwen 8B adapter pools via LightGBM.
    \item The PAS response architecture enforcing structural evidence assignment to improve explanation reliability.
\end{itemize}

\section{Method}

As shown in Figure~\ref{fig:pipeline}, our pipeline sequentially enriches the track catalog with external metadata, routes dialogue through two complementary retrieval pools, calibrates candidate rankings via LightGBM, and structures the final output using an evidence-grounded PAS framework.

\subsection{Official Data and External Metadata}

The official TalkPlayData-Challenge dataset suite\footnote{\url{https://huggingface.co/datasets/talkpl-ai/TalkPlayData-Challenge-Dataset}} supplies a 47,071-track catalog along with dialogue sessions and conversation goals. To mitigate data sparsity, we integrate external metadata from TalkPlayTools\allowbreak-Env\footnote{\url{https://huggingface.co/datasets/talkpl-ai/TalkPlayTools-Env}} for descriptive audio, LRCLIB\footnote{\url{https://lrclib.net/}} for lyrics fallback, and MusicBrainz\footnote{\url{https://musicbrainz.org/}} for structured fields. As shown in Table~\ref{tab:metadata}, these sources provide complementary coverage, enriching 96.6\% of the catalog with at least one external field.

\begin{table}[h]
  \caption{Coverage of external metadata over the 47,071-track catalog.
  Source rows may overlap.}
  \label{tab:metadata}
  \centering
  \scriptsize
  \renewcommand{\arraystretch}{0.9}
  \setlength{\tabcolsep}{4pt}
  \begin{tabular}{@{}lp{3.3cm}rr@{}}
    \toprule
    Source & Information used & Tracks & Coverage \\
    \midrule
    TalkPlayTools-Env & lyrics, caption, chord, tempo, key & 13,745 & 29.2\% \\
    LRCLIB & lyrics fallback & 28,162 & 59.8\% \\
    MusicBrainz & IDs, release, label, country, tags & 38,585 & 82.0\% \\
    Any source & at least one sourced field & 45,475 & 96.6\% \\
    \bottomrule
  \end{tabular}
  \vspace{-2mm}
\end{table}

For example, the track ``Dollhouse'' in Figure~\ref{fig:textviews}b expands its baseline gothic and industrial tags with external fields such as 130.21 BPM and D\# minor. This enriched profile flows downstream: keywords expand BM25 documents, structured properties populate conversational track briefs, and captions and lyrics ground PAS response generation; missing fields are strictly omitted.

\subsection{Pool-Specific Retrieval Interfaces}

\begin{figure*}[t]
  \centering
  {\sffamily\scriptsize\bfseries (a) Final pool topology}\par\vspace{1pt}
  \resizebox{0.95\textwidth}{!}{%
  \begin{tikzpicture}[node distance=3.5mm and 4.5mm]
    \node[figinput,align=center,text width=18mm,inner sep=2.5pt] (hybridin)
      {Dialogue history\\+ track catalog};
    \node[figinterface,align=left,text width=59mm,inner sep=2.5pt,right=of hybridin] (hybridlogic)
      {\textbf{BM25:} catalog + external keywords $\rightarrow$ two searches $\rightarrow$ RRF\\
       \textbf{0.6B dense:} catalog strings $\rightarrow$ embedding search};
    \node[figprocess,align=center,text width=19mm,inner sep=2.5pt,right=of hybridlogic] (hybridrrf)
      {cross-component\\RRF};
    \node[figoutput,align=center,text width=27mm,inner sep=2.5pt,right=of hybridrrf] (hybridout)
      {\textbf{Hybrid pool}\\\texttt{bm25\_qmr}};

    \node[figinput,align=center,text width=18mm,inner sep=2.5pt,below=of hybridin] (convin)
      {Dialogue history\\+ track catalog};
    \node[figinterface,align=center,text width=34mm,inner sep=2.5pt,right=of convin] (convlogic)
      {8B structured query\\+ track brief};
    \node[figprocess,align=center,text width=27mm,inner sep=2.5pt,right=of convlogic] (foldsearch)
      {five adapted 8B searches};
    \node[figprocess,align=center,text width=16mm,inner sep=2.5pt,right=of foldsearch] (foldrrf)
      {five-fold\\RRF};
    \node[figoutput,align=center,text width=31mm,inner sep=2.5pt,right=of foldrrf] (convout)
      {\textbf{Conversational pool}\\\texttt{qemb\_twotower\_8b}};

    \draw[-{Latex[length=2mm]},draw=FigNavy,line width=0.9pt] (hybridin.east) -- (hybridlogic.west);
    \draw[-{Latex[length=2mm]},draw=FigNavy,line width=0.9pt] (hybridlogic.east) -- (hybridrrf.west);
    \draw[-{Latex[length=2mm]},draw=FigNavy,line width=0.9pt] (hybridrrf.east) -- (hybridout.west);
    \draw[-{Latex[length=2mm]},draw=FigNavy,line width=0.9pt] (convin.east) -- (convlogic.west);
    \draw[-{Latex[length=2mm]},draw=FigNavy,line width=0.9pt] (convlogic.east) -- (foldsearch.west);
    \draw[-{Latex[length=2mm]},draw=FigNavy,line width=0.9pt] (foldsearch.east) -- (foldrrf.west);
    \draw[-{Latex[length=2mm]},draw=FigNavy,line width=0.9pt] (foldrrf.east) -- (convout.west);
  \end{tikzpicture}}

  \vspace{2pt}
  {\sffamily\scriptsize\bfseries (b) Shortened turn-6 query/document views with temporal alignment}\par\vspace{2pt}
  \tiny
  \setlength{\tabcolsep}{0pt}
  \renewcommand{\arraystretch}{1.0}
  
  \begin{tabular*}{\linewidth}{@{\extracolsep{\fill}}lp{0.36\linewidth}p{0.34\linewidth}p{0.14\linewidth}@{}}
    \toprule
    \textbf{Component} & \textbf{Query-side text (Inputs)} & \textbf{Document-side text (Target)} & \textbf{Preserved signal} \\
    \midrule
    \textbf{Hybrid: BM25} &
    \textbf{[Turns 1--5]} \texttt{Burn Witch Burn; Ego Likeness; 2006} \newline 
    \textbf{[Turn 6]} \texttt{gothic rock; female vocals; industrial; driving beat} &
    \texttt{Dollhouse (Razed In Black Mix); Switchblade Symphony; 2001}; tags: \texttt{gothic, darkwave, industrial}. & 
    Exact entities and current search tags. \\
    \midrule 
    \textbf{Hybrid: 0.6B dense} &
    \textbf{[Turns 1--5]} \texttt{title: burn witch burn, artist: ego likeness} \newline 
    \textbf{[Turn 6]} Lowercase text-based search tokens. & 
    \texttt{Dollhouse; Switchblade Symphony; 2001; popularity 11}; tags: \texttt{punk, ebm, remix}. & 
    Catalog semantics; official fields only. \\
    \midrule
    \textbf{Conversational: 8B} &
    Fixed four-field structured prompt template. \newline (Instantiated in the table below.) & 
    \texttt{Dollhouse} by Switchblade Symphony; 130 BPM, D\# minor, gothic/darkwave. & 
    Evolving conversational intent across turns. \\
    \bottomrule
  \end{tabular*}
  


  \begin{tabular*}{\linewidth}{@{\extracolsep{\fill}} p{0.22\linewidth} p{0.74\linewidth} @{}}
    \toprule
    \multicolumn{2}{@{}l}{\textbf{Actual 8B Structured Query Fields for Turn-6 Execution}} \\
    \midrule
    \textbf{History [Turns 1--5]} & 
    \textbf{Recent listens:} Burn Witch Burn [T5]; In The Throat Of The Unbounded; Sorrow Is Her Name; Quiet Moments; Frozen. \newline
    \textbf{Session so far:} User anchored in Lycia's ethereal darkwave, expanding through Violet Tears and Ego Likeness toward gothic rock. \\
    \midrule
    \textbf{Current [Turn 6]} & 
    \textbf{Looking for:} gothic rock + industrial elements + heavier driving beat. \newline
    \textbf{Context:} female vocals; Switchblade Symphony; gothic/darkwave. \\
    \bottomrule
  \end{tabular*}
  \caption{Two final-pool pipelines and temporal separation of inputs. (a) Topology of the hybrid and conversational pipelines. (b) Input-target pairs for Turn 6. The bottom subtable details the 8B model's query structure.}
  \Description{Panel a shows the hybrid pipeline, where BM25 and a 0.6B dense component merge through cross-component RRF into the bm25 qmr pool, and the conversational pipeline, where five adapted 8B searches merge through five-fold RRF into the qemb twotower 8B pool. Panel b lists the turn-6 query-side and document-side text for each component and the structured 8B query fields.}
  \label{fig:textviews}
\end{figure*}
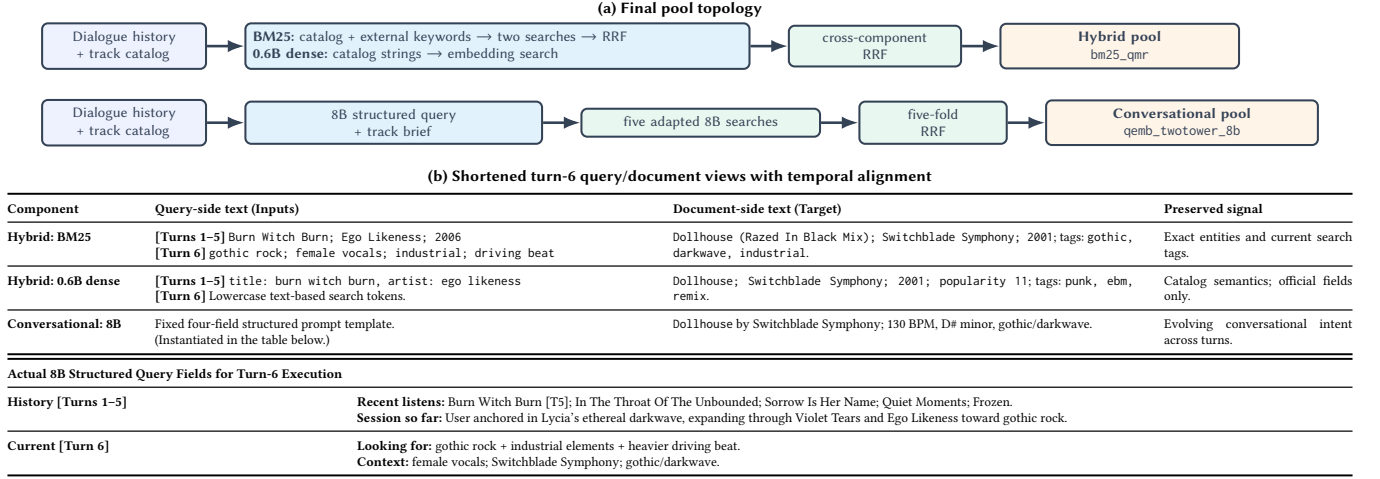

Rather than issuing a single query, our architecture routes retrieval through two parallel interfaces (Figure~\ref{fig:textviews}). The hybrid pool (\texttt{bm25\_qmr}) combines sparse and dense mechanisms. The sparse component runs two BM25 searches~\cite{robertson1994okapi}: a baseline catalog search and a variant enriched with catalog tags and external keywords. Its query fuses metadata from prior music turns with filtered current-request tokens, and an RRF layer~\cite{cormack2009rrf} merges their rankings. The dense component uses a 0.6B Qwen3 embedding model~\cite{qwen3embedding2025}, joining historical metadata with the lowercased current request to query official document vectors. A second RRF fuses the sparse and dense rankings without score calibration, producing the hybrid pool's candidate list.

The conversational pool (\texttt{qemb\_twotower\_8b}) relies entirely on natural language on both sides to track long-range session dynamics. The query format strictly structures the dialogue history into four ordered fields: \texttt{Recent listens:} lists prior tracks chronologically; \texttt{Session so far:} supplies an optional cached preference trajectory summary under 25 words; \texttt{Context:} aggregates explicit goals, eras, and popularity cues; and \texttt{Looking for:} contains the current explicit request. When context window limits are reached, a strict truncation priority rule preserves the user's immediate intent: because \texttt{Looking for:} sits adjacent to the final embedding token, it is never dropped, whereas lower-priority historical context fields are truncated first. To optimize this contextual mapping, we fine-tune an 8B parameter encoder model using 4-bit QLoRA~\cite{dettmers2023qlora} under an in-batch InfoNCE contrastive objective~\cite{oord2018cpc} with LogQ frequency correction~\cite{yi2019sampling}. The dense retrieval is executed across five session-level cross-validation folds, and their rankings are combined via RRF into the unified \texttt{qemb\_twotower\_8b} pool (bagging).\looseness=-1

Turn 1 lacks dialogue history: BM25 and dense queries collapse into the current request tokens, and the 8B query retains only the \texttt{Context:} and \texttt{Looking for:} fields. Document representations remain unchanged, and the downstream LightGBM feature row records the turn number.

\subsection{Candidate Fusion and Calibration}

To consolidate retrieval, the system unions the top-100 candidates from each pool. A LightGBM~\cite{ke2017lightgbm} ranker then scores and calibrates this final list. Table~\ref{tab:features} outlines the input feature schema. The model ingests individual ranks and scores from both pools, alongside their cross-pool RRF metrics. Turn number, request category, and goal specificity define the session context. Non-retrieved candidates receive sentinel values. Training uses out-of-fold scoring to prevent data leakage from the 8B retrieval model.

\begin{table}[h]
  \caption{LightGBM feature schema.}
  \label{tab:features}
  \centering
  \scriptsize
  \renewcommand{\arraystretch}{0.9}
  \setlength{\tabcolsep}{5pt}
  \begin{tabular}{@{}lll@{}}
    \toprule
    Source & Fields & Decision evidence \\
    \midrule
    Hybrid pool & rank, score & exact/general support \\
    Conversational pool & rank, score & dialogue support \\
    Cross-pool RRF & rank, score & retrieval agreement \\
    Context & turn, category, specificity & request regime \\
    \bottomrule
  \end{tabular}
\end{table}

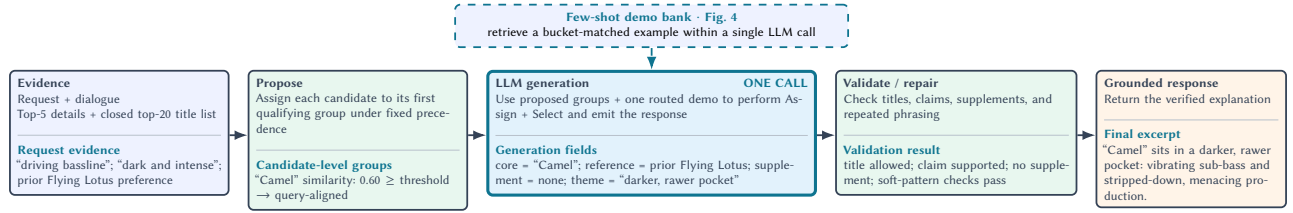
\begin{figure*}[t]
  \centering
  \resizebox{0.95\textwidth}{!}{%
  \begin{tikzpicture}[node distance=3mm,figure-readable]
    \node[figinput,align=left,text width=28mm,inner sep=3pt,font=\sffamily\tiny] (ev)
      {\textbf{Evidence}\par
       Request + dialogue\par
       Top-5 details + closed top-20 title list\par
       \vspace{1pt}\textcolor{FigLine}{\rule{\linewidth}{0.4pt}}\par\vspace{1pt}
       \textcolor{FigTab}{\bfseries Request evidence}\par
       ``driving bassline''; ``dark and intense''; prior Flying Lotus preference};
    \node[figprocess,align=left,text width=28mm,inner sep=3pt,font=\sffamily\tiny,
      right=2.5mm of ev.north east,anchor=north west] (prop)
      {\textbf{Propose}\par
       Assign each candidate to its first qualifying group under fixed precedence\par
       \vspace{1pt}\textcolor{FigLine}{\rule{\linewidth}{0.4pt}}\par\vspace{1pt}
       \textcolor{FigTab}{\bfseries Candidate-level groups}\par
       ``Camel'' similarity: 0.60 $\geq$ threshold\par
       $\rightarrow$ query-aligned};
    \node[figfocus,align=left,text width=43mm,inner sep=3pt,font=\sffamily\tiny,
      right=2.5mm of prop.north east,anchor=north west] (llm)
      {\textbf{LLM generation}\hfill\textcolor{FigTab}{\bfseries ONE CALL}\par
       Use proposed groups + one routed demo to perform Assign + Select and emit
       the response\par
       \vspace{1pt}\textcolor{FigLine}{\rule{\linewidth}{0.4pt}}\par\vspace{1pt}
       \textcolor{FigTab}{\bfseries Generation fields}\par
       core = ``Camel''; reference = prior Flying Lotus; supplement = none;
       theme = ``darker, rawer pocket''};
    \node[figprocess,align=left,text width=31mm,inner sep=3pt,font=\sffamily\tiny,
      right=2.5mm of llm.north east,anchor=north west] (sel)
      {\textbf{Validate / repair}\par
       Check titles, claims, supplements, and repeated phrasing\par
       \vspace{1pt}\textcolor{FigLine}{\rule{\linewidth}{0.4pt}}\par\vspace{1pt}
       \textcolor{FigTab}{\bfseries Validation result}\par
       title allowed; claim supported; no supplement; soft-pattern checks pass};
    \node[figoutput,align=left,text width=24mm,inner sep=3pt,font=\sffamily\tiny,
      right=2.5mm of sel.north east,anchor=north west] (resp)
      {\textbf{Grounded response}\par
       Return the verified explanation\par
       \vspace{1pt}\textcolor{FigLine}{\rule{\linewidth}{0.4pt}}\par\vspace{1pt}
       \textcolor{FigTab}{\bfseries Final excerpt}\par
       ``Camel'' sits in a darker, rawer pocket: vibrating sub-bass and
       stripped-down, menacing production.};
    \draw[figarrow] ([yshift=-9mm]ev.north east)--([yshift=-9mm]prop.north west);
    \draw[figarrow] ([yshift=-9mm]prop.north east)--([yshift=-9mm]llm.north west);
    \draw[figarrow] ([yshift=-9mm]llm.north east)--([yshift=-9mm]sel.north west);
    \draw[figarrow] ([yshift=-9mm]sel.north east)--([yshift=-9mm]resp.north west);

    \node[draw=FigTab,dashed,line width=0.7pt,fill=FigLavender,rounded corners=2pt,
      align=center,text width=52mm,inner sep=3pt,font=\sffamily\tiny,
      above=3mm of llm] (demo)
      {\textcolor{FigTab}{\bfseries Few-shot demo bank $\cdot$ Fig.~\ref{fig:demo-compiler}}\par
       retrieve a bucket-matched example within a single LLM call};
    \draw[figdash] (demo.south)--(llm.north);
  \end{tikzpicture}}
  \caption{PAS runtime execution mapped to a single-shot demonstration workflow. Candidate-level evidence groups and a bucket-matched example condition a single LLM call, followed by deterministic validation.}
  \Description{Five stages flow left to right: evidence collection from the request and dialogue, proposal of candidate-level evidence groups, a single LLM generation call conditioned on a routed demonstration from the few-shot demo bank, deterministic validation and repair, and the final grounded response.}
  \label{fig:pas}
  \vspace{-1mm}
\end{figure*}

\begin{figure*}[t]
  \centering
  \resizebox{0.95\textwidth}{!}{%
  \begin{tikzpicture}[node distance=3mm,figure-readable]
    \node[figinput,align=left,text width=22mm,inner sep=2.5pt,font=\sffamily\tiny] (turns)
      {\textbf{Eligible training turns}\par
       Sample completed recommendation turns\par
       \vspace{1pt}\textcolor{FigLine}{\rule{\linewidth}{0.4pt}}\par\vspace{1pt}
       \textcolor{FigTab}{\bfseries Compiler input}\par
       1,000 turns, not sessions};
    \node[figinterface,align=left,text width=26mm,inner sep=2.5pt,font=\sffamily\tiny,
      right=2.5mm of turns.north east,anchor=north west] (judge)
      {\textbf{Generate + judge}\par
       GLM-5.2 candidate generation\par
       Personalization, Explanation, and lexical novelty\par
       \vspace{1pt}\textcolor{FigLine}{\rule{\linewidth}{0.4pt}}\par\vspace{1pt}
       \textcolor{FigTab}{\bfseries Quality gates}\par
       15--130 words; total score $\geq$ 0.7};
    \node[figprocess,align=left,text width=24mm,inner sep=2.5pt,font=\sffamily\tiny,
      right=2.5mm of judge.north east,anchor=north west] (coverage)
      {\textbf{Coverage selection}\par
       Balance category, specificity, and turn position\par
       \vspace{1pt}\textcolor{FigLine}{\rule{\linewidth}{0.4pt}}\par\vspace{1pt}
       \textcolor{FigTab}{\bfseries Selected bank}\par
       18 demonstrations};
    \node[figfocus,align=left,text width=32mm,inner sep=2.5pt,font=\sffamily\tiny,
      right=2.5mm of coverage.north east,anchor=north west] (bank)
      {\textbf{Bucketed demo bank}\par
       Index by category $\times$ specificity $\times$ turn\par
       \vspace{1pt}\textcolor{FigLine}{\rule{\linewidth}{0.4pt}}\par\vspace{1pt}
       \textcolor{FigTab}{\bfseries Matched bucket}\par
       key by A | HL | mid\par
       \textcolor{FigMuted}{compiled once; reused at runtime}};
    \node[figprocess,align=left,text width=25mm,inner sep=2.5pt,font=\sffamily\tiny,
      right=2.5mm of bank.north east,anchor=north west] (retrieve)
      {\textbf{Retrieve one demo}\par
       Select one compiled response example from the matched bucket\par
       \vspace{1pt}\textcolor{FigLine}{\rule{\linewidth}{0.4pt}}\par\vspace{1pt}
       \textcolor{FigTab}{\bfseries Selected example}\par
       core track: ``Camel'' by Flying Lotus};
    \node[figoutput,draw=FigTab,line width=1.0pt,align=left,text width=22mm,inner sep=2.5pt,
      font=\sffamily\tiny,right=2.5mm of retrieve.north east,anchor=north west] (handoff)
      {\textbf{Condition Fig.~\ref{fig:pas}}\par
       Insert bucket-matched example as one-shot response guidance
       \vspace{1pt}\textcolor{FigLine}{\rule{\linewidth}{0.4pt}}\par\vspace{1pt}
       \textcolor{FigTab}{\bfseries Destination}\par
       single LLM generation call};

    \draw[figarrow] ([yshift=-9mm]turns.north east)--([yshift=-9mm]judge.north west);
    \draw[figarrow] ([yshift=-9mm]judge.north east)--([yshift=-9mm]coverage.north west);
    \draw[figarrow] ([yshift=-9mm]coverage.north east)--([yshift=-9mm]bank.north west);
    \draw[figarrow] ([yshift=-9mm]bank.north east)--([yshift=-9mm]retrieve.north west);
    \draw[figarrow] ([yshift=-9mm]retrieve.north east)--([yshift=-9mm]handoff.north west);

    \draw[draw=FigLine,line width=0.7pt]
      ([yshift=2mm]turns.north west)--([yshift=2mm]bank.north east);
    \node[font=\sffamily\tiny\bfseries,text=FigMuted,fill=white,inner sep=1pt,
      above=2mm of $(turns.north)!0.5!(bank.north)$] {COMPILE ONCE OFFLINE};
    \draw[draw=FigTab,line width=0.7pt]
      ([yshift=2mm]retrieve.north west)--([yshift=2mm]handoff.north east);
    \node[font=\sffamily\tiny\bfseries,text=FigTab,fill=white,inner sep=1pt,
      above=2mm of $(retrieve.north)!0.5!(handoff.north)$] {RETRIEVE AT RUNTIME};

    \node[draw=FigTab,dashed,line width=0.7pt,fill=FigLavender,rounded corners=2pt,
      align=center,text width=32mm,inner sep=2.5pt,font=\sffamily\tiny,
      below=3mm of bank] (request)
      {\textcolor{FigTab}{\bfseries Current request signature}\par category =  \texttt{A} | specificity = \texttt{HL} | turn = \texttt{mid}};
    \draw[figdash] (request.north)--(bank.south);
  \end{tikzpicture}}
  \caption{Offline demonstration compilation and runtime routing mechanisms. The current request signature queries the shared 18-example bank. The \texttt{A|HL|mid} route then selects a matching example.}
  \Description{An offline compile-once stage samples eligible training turns, generates and judges candidate responses, applies coverage selection, and stores 18 demonstrations in a bucketed bank. At runtime the current request signature retrieves one bucket-matched demonstration, which conditions the single LLM generation call of the PAS figure.}
  \label{fig:demo-compiler}
  \vspace{-1mm}
\end{figure*}
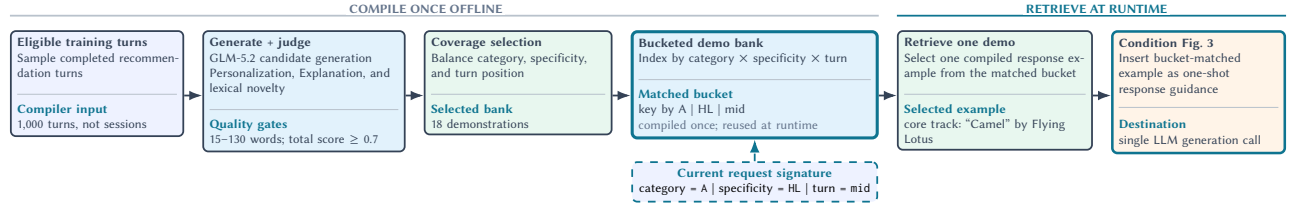

\begin{table*}[!t]
  \begin{minipage}[t]{0.49\textwidth}
    \caption{Public-test retrieval results; 0.6B is not task-adapted.}
    \label{tab:retrieval}
    \centering
    \scriptsize
    \renewcommand{\arraystretch}{0.9}
    \setlength{\tabcolsep}{5pt}
    \begin{tabular}{@{}lccc@{}}
      \toprule
      Configuration & nDCG & R@20 & R@100 \\
      \midrule
      Enriched BM25 ensemble & .1599 & .3321 & .4793 \\
      Untrained 0.6B dense & .1210 & .2385 & .3531 \\
      Hybrid lexical--dense & .1550 & .3170 & .4639 \\
      8B five-fold RRF & .2165 & .4363 & .6619 \\
      Full cross-pool RRF & .2198 & \textbf{.4406} & \textbf{.6630} \\
      + LightGBM & \textbf{.2208} & .4400 & \textbf{.6630} \\
      \bottomrule
    \end{tabular}
  \end{minipage}\hfill
  \begin{minipage}[t]{0.47\textwidth}
    \caption{Blind-A response check on fixed track IDs.}
    \label{tab:response}
    \centering
    \scriptsize
    \renewcommand{\arraystretch}{0.9}
    \setlength{\tabcolsep}{5pt}
    \begin{tabular}{@{}lccc@{}}
      \toprule
      Generator & Demos & Judge & Composite \\
      \midrule
      GLM-5.2 PAS & 3 & 4.80 & .6556 \\
      Gemini-3.1-Flash-Lite PAS & 3 & 4.75 & .6518 \\
      GLM-5.2 PAS & 1 & \textbf{4.90} & \textbf{.6631} \\
      \bottomrule
    \end{tabular}
  \end{minipage}
\end{table*}

\begin{table}[!t]
  \caption{Category nDCG@20 and relative gains (\%) by retrieval pool and
  dialogue stage. Bold marks the best retrieval result in each category.}
  \label{tab:pool-category}
  \centering
  \fontsize{6}{6.9}\selectfont
  \setlength{\tabcolsep}{1.7pt}
  \resizebox{\columnwidth}{!}{%
  \begin{tabular}{@{}lrrrrrrrrr@{}}
    \toprule
    Category (N) & BM25 & 0.6B & Adapt. 8B & Fusion & $\Delta_{\mathrm{crawl}}$ & $\Delta_{\mathrm{best}}$ & Early & Mid. & Late \\
    \midrule
    Sound traits (61) & .161 & .147 & \textbf{.219} & .216 & $+5.5$ & $-1.7$ & $-7.2$ & $-0.4$ & $+0.8$ \\
    Lyrics (142) & .178 & .126 & .224 & \textbf{.229} & $+6.3$ & $+2.5$ & $+5.5$ & $+4.7$ & $-1.8$ \\
    Artwork (58) & .139 & .109 & .174 & \textbf{.176} & $+2.7$ & $+0.8$ & $-0.8$ & $-3.2$ & $+7.5$ \\
    Activity fit (86) & .152 & .119 & .200 & \textbf{.205} & $+2.9$ & $+2.3$ & $+3.8$ & $-2.4$ & $+7.2$ \\
    Discovery (95) & .173 & .127 & .214 & \textbf{.217} & $+4.9$ & $+1.1$ & $-0.6$ & $-1.0$ & $+5.0$ \\
    Memory (95) & .193 & .153 & .275 & \textbf{.281} & $+3.2$ & $+2.1$ & $+4.0$ & $+2.3$ & $+0.2$ \\
    Mood (77) & .156 & .098 & .197 & \textbf{.201} & $+7.7$ & $+1.7$ & $-2.3$ & $+1.4$ & $+5.2$ \\
    Artist (135) & .177 & .150 & .242 & \textbf{.245} & $+2.4$ & $+1.2$ & $+2.2$ & $-3.1$ & $+5.2$ \\
    Exact hit (18) & .118 & .070 & .164 & \textbf{.180} & $+24.4$ & $+10.1$ & $+9.3$ & $+14.7$ & $+6.5$ \\
    Trends (77) & .136 & .095 & .211 & \textbf{.217} & $+2.9$ & $+3.2$ & $-1.6$ & $+1.1$ & $+11.6$ \\
    Era (156) & .132 & .094 & \textbf{.196} & .195 & $+7.9$ & $-0.1$ & $-3.0$ & $+1.5$ & $+1.2$ \\
    \bottomrule
  \end{tabular}}
\end{table}

\subsection{PAS Response Generation}
\label{sec:pas_generation}

We adapt the PAS framework~\cite{wang2023goalex} to enforce structural constraints on response claims for the top-20 ranked tracks. Figure~\ref{fig:pas} illustrates the complete runtime execution alongside a sample trace. First, the propose stage partitions the 20 candidates into distinct evidence groups under fixed precedence rules: query alignment, lyric resonance, taste continuity, pool coherence, and discovery. In the sample trace, tokens matching ``driving bassline'' and ``dark and intense'' yield a 0.60 similarity score for ``Camel'' by Flying Lotus, satisfying the query-alignment threshold.

A single structured LLM call processes the proposed candidate groups together with one routed demonstration. This demonstration provides response selection guidance rather than introducing additional recommendation candidates. Second, within this single call, the assign step identifies usable personalization anchors and supported themes. Third, the select step chooses a core track, optional supplementary tracks for requests seeking multiple tracks, and optional historical references. The same joint call generates the raw response text. Lastly, deterministic rules validate and repair title formatting errors or citation violations. If the draft matches any blacklisted session phrases, PAS triggers a single generation retry.\looseness=-1

Figure~\ref{fig:demo-compiler} outlines the offline compilation process for the reusable demonstration bank. The compilation pipeline samples 1,000 eligible recommendation turns from the TalkPlayData training split. GLM-5.2 generates the initial candidate responses. An independent LLM-as-judge scores personalization and explanation metrics on a 1-to-5 scale. We normalize these outputs into a unified quality metric. The final score weights this quality metric at 75\% and lexical novelty at 25\%. Viable candidates must contain between 15 and 130 words and achieve a minimum combined score of 0.7. A coverage-aware filtering step balances the final selection across request category, specificity, and turn position. This selection process yields a fixed 18-example runtime demonstration bank.\looseness=-1

At runtime, each incoming request derives a routing key based on its category, specificity, and turn. The router evaluates the exact matching bucket first. If that bucket is empty, the system sequentially falls back to category matches, specificity or turn matches, and the global bank. The router injects the highest-scoring demonstration from the first nonempty layer. Figure~\ref{fig:demo-compiler} displays the key \texttt{A|HL|mid}, where \texttt{A} represents a sound-focused category, \texttt{HL} encodes high goal specificity with low track specificity, and \texttt{mid} indicates a mid-dialogue turn. The trace shown in Figure~\ref{fig:pas} instantiates this routed example.\looseness=-1

\section{Experiments}
\label{sec:experiments}

\subsection{Evaluation Setup}

We score 8,000 turns from the 1,000-session public test split against all 47,071
tracks using nDCG@20, R@20, and R@100. Blind-A and Blind-B contain 80 truncated
sessions each and are used only for final external checks. Response ablations
fix track IDs. In Table~\ref{tab:response}, Judge is the score returned by the
organizers' undisclosed LLM judge, and Composite is the official aggregate
score.

\subsection{Results and Controlled Ablations}

Table~\ref{tab:retrieval} summarizes public-test retrieval results. Individual 8B folds
range from .2018 to .2138 in nDCG@20, from .4100 to .4298 in R@20, and from
.6369 to .6435 in R@100. Five-fold RRF provides the largest ranking gain. It improves
over the mean individual fold by .0076 nDCG@20 and raises R@100 from .6409 to
.6619.

The pools recover different targets. At least one pool places the target in the
top 20 on 3,838 turns. Both succeed on 2,188; the conversational pool alone
succeeds on 1,302, and the hybrid pool alone succeeds on 348. Their top-20
candidate sets have a mean Jaccard overlap of .231. This complementarity makes
fusion useful: cross-pool RRF raises nDCG@20 from .2165 to .2198. LightGBM
further raises it to .2208, while R@20 changes from .4406 to .4400 and R@100
remains .6630. These mixed changes support interpreting LightGBM as
candidate-order calibration rather than as a broad retrieval gain.

With tracks fixed, Table~\ref{tab:response} isolates response generation. Among
GLM-5.2 runs, one routed demonstration outperforms three. On Blind-B, PAS raises the
official composite from .55 to .58. This supports the routed one-shot choice
under the automatic judge but does not support a claim about human preference.

\subsection{Pool Behavior by Request and Turn}

Table~\ref{tab:pool-category} tests whether fusion helps uniformly. For each
subset, $\Delta_{\mathrm{best}}$ is $100(\mathrm{Fusion}/\mathrm{best}-1)$, where
best is the strongest of BM25, 0.6B, and 8B. $\Delta_{\mathrm{crawl}}$ is
$100(\mathrm{BM25}/\mathrm{BM25}_{\mathrm{no\ crawl}}-1)$. Early, Mid., and
Late report $\Delta_{\mathrm{best}}$ for turns 1--2, 3--5, and 6--8; $N$ is
the number of sessions.

Bold values in Table~\ref{tab:pool-category} identify each category's best
result. The adapted 8B pool is the strongest individual pool in all eleven
categories; fusion is best in nine. Its benefit appears
more often later in the dialogue: fusion improves over the best individual pool
in five categories during early turns, six during middle turns, and ten during
late turns.
Sound-focused and era-focused requests are the two aggregate exceptions.

Crawl-derived fields raise enriched BM25 from .1524 to .1599 nDCG@20, an aggregate
4.9\% gain, with category-level gains ranging from +2.4\% to +24.4\%
(Table~\ref{tab:pool-category}). The untrained 0.6B
dense component does not improve BM25 on the public test split; adding it
lowers the hybrid result from .1599 to .1550. Replacing BM25's music-only query
with the full dialogue further lowers hybrid nDCG@20 to .1527. The released
BM25 interface therefore retains prior track metadata and the current request
but excludes earlier user and assistant text.

\section{Analysis and Limitations}

Three design lessons emerge. Tailoring input text to each scoring mechanism lets the hybrid interface exploit exact catalog cues while the conversational interface tracks evolving language, as Figure~\ref{fig:textviews} shows. The low candidate overlap and complementary target hits between these channels create productive pool disagreement, explaining why cross-pool RRF outperforms single-pool baselines (Section~\ref{sec:experiments}). Finally, explicit evidence and role assignments let PAS prioritize evidence-backed selection over surface realization.\looseness=-1

These findings point toward extending the LightGBM ranker, which currently relies on only two upstream pools and compact features; stronger retrieval components and richer track--dialogue features would give the calibrator more informative signals and further extend its reranking gains.\looseness=-1

Several limitations remain. Uneven external metadata can constrain PAS despite deterministic validation, and the in-house judge may introduce model bias during offline compilation. The 18-example synthetic bank also cannot cover every dialogue state or sparse request bucket, and the five adapted 8B folds plus LLM-based synthesis require substantial compute. This study therefore leaves runtime latency, serving cost, and human preference unevaluated.\looseness=-1

\section{Conclusion}

Isolating retrieval from response generation optimizes recommendation accuracy and system transparency. Pool-specific text interfaces safeguard exact catalog cues, cross-pool RRF captures complementary ranking signals, and the PAS framework prevents ungrounded claims through structured constraints. Empirical evaluation confirms that external crawl enrichment drives performance gains across all request categories, while fusion delivers its largest benefits in late-dialogue turns. This clear separation of concerns supported our third-place finish in the Blind-B industry track, proving that decoupling retrieval from surface realization maintains competitive performance while keeping all decisions fully inspectable.\looseness=-1

\clearpage
\begingroup
\interlinepenalty=10000
\bibliographystyle{ACM-Reference-Format}
\bibliography{references}
\endgroup

\ifdefined\withappendix
  \clearpage
  \input{appendix}
\fi

\end{document}